%% file: conference_101719.tex
\documentclass[conference]{IEEEtran}
\IEEEoverridecommandlockouts
\usepackage{cite}
\usepackage{amsmath,amssymb,amsfonts}
\usepackage{algorithmic}
\usepackage{algorithm}
\usepackage{graphicx}
\usepackage{textcomp}
\usepackage{xcolor}
\usepackage{url}
\usepackage{array}
\usepackage{enumerate}
\usepackage{verbatim}
\usepackage{xspace}

\def\eg{\emph{e.g. }\xspace}

\def\ie{\emph{i.e.,}\xspace}

\newcommand{\pb}[1]{\vspace{0.75ex}\noindent{\bf \em #1}\hspace*{.3em}}

\newcommand{\zhenyu}[1]{\textcolor{blue}{ZY: #1}}
\newcommand{\penglai}[1]{\textcolor{purple}{PL: #1}}

\def\BibTeX{{\rm B\kern-.05em{\sc i\kern-.025em b}\kern-.08em
    T\kern-.1667em\lower.7ex\hbox{E}\kern-.125emX}}
\begin{document}

\title{NetFC: Enabling Accurate Floating-point Arithmetic on Programmable Switches}
\author{
\vspace{0.2em}
{\rm Penglai Cui$^\dag$, Heng Pan$^\dag$ $^\ddag$, Zhenyu Li$^\dag$ $^\ddag$, Jiaoren Wu$^\S$}\\
\vspace{1.1em}
{\rm Shengzhuo Zhang$^\S$, Xingwu Yang$^\S$, Hongtao Guan$^\dag$,\vspace{-0.8em} Gaogang Xie$^*$}\\
\vspace{-0.8em}
{\normalsize{$^\dag$ICT, CAS, China ~~ $^\ddag$Purple Mountain Laboratories ~~ $^\S$Kuaishou ~~ $^*$CNIC, CAS, China}}}

\maketitle

\begin{abstract}
In-network computation has been widely used to accelerate data-intensive distributed applications. Some computational tasks, traditional performed on servers, are offloaded to the network on programmable switches. However, the computational capacity of programmable switches is limited to simple integer arithmetic operations while many of applications require on-the-fly floating-point operations. To address this issue, prior 
approaches either adopt a float-to-integer method or directly offload computational tasks to the local CPUs of switches, incurring accuracy loss and delayed processing. 

To this end, we propose NetFC, a table-lookup method to achieve on-the-fly in-network floating-point arithmetic operations nearly without accuracy loss. NetFC adopts a divide-and-conquer mechanism that converts the original huge table into several much small tables together with some integer operations.
NetFC further leverages a scaling-factor mechanism for computational accuracy improvement, and a prefix-based lossless table compression method to reduce the memory consumption. We use both synthetic and real-life datasets to evaluate NetFC. The experimental results show that the average accuracy of NetFC is as high as up to 99.94\% at worst with only 448KB memory consumption. Furthermore, we integrate NetFC into Sonata~\cite{gupta2018sonata} for detecting Slowloris attack, yielding significant decrease of detection delay. 


\end{abstract}


\input{introduction/introduction2}
\input{motivation/motivation}

\input{design/design}

\input{implementation/implementation}

\input{evaluation/evaluation}
\input{usecase/usecase}
\input{related/related}
\input{conclusion/conclusion}

\bibliographystyle{plain}
\bibliography{reference}
\end{document}

%% file: introduction/introduction2.tex
\section{Introduction}
In modern data center, many applications are data-intensive, such as big data analysis~\cite{dean2008mapreduce}, distributed deep learning~\cite{peng2019generic}, graph processing~\cite{malewicz2010pregel} and real-time stream processing~\cite{jalaparti2013speeding,castro2013integrating}. Due to frequent data exchanging, these applications have suffered performance degradation from network overhead. For example, in some of the FaceBook MapReduce jobs, network communication can occupy up to 70\% of the execution time~\cite{chowdhury2011managing}; in distributed reinforcement learning, network overhead can be over 80\% of the total cost in each iteration~\cite{li2019accelerating}. Thus cutting down network traffic to accelerate network communication has been the key factor to improve the overall performance.

Recently, a new direction to accelerate data-intensive applications, \emph{i.e.} in-network computation, has been explored. The intuition behind is that the network has been equipped with many new devices (e.g. programmable switches~\cite{tofino} and smart network interface~\cite{le2017uno}); that said the network has become capable of providing computational capacity. Consequently, some computational tasks, traditionally performed in the host side, can be offloaded to the network devices. And network traffic can be intercepted and processed by the network devices on the fly based on the pre-offloaded computational logics before it reaches the hosts. Indeed, researchers in the community have already utilized in-network computation to accelerate different distributed applications and achieve remarkable performance improvement. For example, NetCache~\cite{jin2017netcache} is co-designed with programmable switches to achieve more than 2B queries per second; ATP~\cite{lao2021atp} performs in-network gradient aggregation, which can improve the training throughput of existing distributed deep learning systems up to 66\%; Linear-Road~\cite{jepsen2018life}, a stream processing benchmark, shows the capability that achieves 4B events per second; Sonata~\cite{gupta2018sonata} utilizes programmable switches to achieve fast In-band network telemetry. 

Unfortunately, the computational capacity of the network is actually very limited, and even the state-of-the-art programmable switches (e.g. Barefoot Tofino~\cite{tofino}) only support simple integer arithmetic operations (e.g. addition and subtraction). Consequently, this becomes a barrier to in-network acceleration of applications, because many of them often require processing sophisticated floating-point data and arithmetic operations (e.g multiplication and division). For example, ATP~\cite{lao2021atp} accelerates distributed deep learning via in-network gradient aggregation, which needs to perform floating-point calculation on programmable switches; Sonata~\cite{gupta2018sonata} measures the state of the network, where some measurement tasks (e.g. Slowloris attack detection~\cite{Slowloris2009}) require in-network multiplication or division. 

To overcome the barrier, the prior works often adopt two different ways of doing this. One is to convert floating-point numbers into integers on the host side in advance so that it only needs to perform integer operations on programmable switches. Indeed, this approach has been adopted by ATP to achieve in-network gradient aggregation. However, it may incur some non-negligible accuracy loss (see Section~\ref{sec:evaluation}). The other solution is to offload the computational tasks to the local CPUs of switches (like Sonata). Nevertheless, this solution introduces significant additional latency (see section~\ref{sec:usecase}). In summary, to the best of our knowledge, there is a lack of a solution to achieve on-the-fly in-network floating-point arithmetic operations nearly without accuracy loss on programmable switches.

To address this gap, we design and implement a novel approach --- NetFC. NetFC adopts a table-lookup method to achieve floating-point arithmetic operations on programmable switches. Intuitively, a simple and direct way is to use a table to enumerate all possible calculation cases ahead. In this way, for one arithmetic operation, we can use its two operands as a key to look up the table while the corresponding value is the result. It is noteworthy that the size of the generated table is huge since it needs to traverse all operands and enumerate their various combinations. For example, for two 16-bit floating-point operands, it would cost almost 8 GB memory, which is unaffordable for on-chip memory on programmable switches (e.g. Barefoot Tofino switches). To this end, NetFC adopts a divide-and-conquer method to address this memory issue. Specifically, it utilizes logarithm projection and transformation to convert the original large table into several much small tables together with simple integer operations (i.e. addition and subtraction). Furthermore, NetFC adopts a scaling-factor mechanism to improve computational accuracy, and design a prefix-based loss-less compression method to further reduce on-chip memory usage. Experimental results on different types of datasets demonstrate that the average accuracy of NetFC is up to 99.94\% at worst with only 448KB memory consumption, which performs much better than the state-of-the-art approach (i.e. float-to-integer). In addition, we integrate NetFC into a network telemetry system---Sonata--- for detecting and defencing Slowloris attack in real time; with NetFC, the detection latency is reduced from 43.16\emph{ms} to 0.046\emph{ms}.

To sum up, the contributions of this paper are three-fold:
\begin{itemize}
  \item We design a table-lookup approach, NetFC, to achieve in-network floating-point arithmetic operations nearly without accuracy loss. It adopts a divide-and-conquer method to address the on-chip memory consumption problem.  

  \item We propose a scaling-factor method to improve the computational accuracy of NetFC, and design a prefix-based loss-less compression mechanism to further reduce memory consumption.

  \item We implement NetFC based on Barefoot Tofino switches. Extensive experiments show that NetFC enables floating-point arithmetic operations  on programmable switches with low on-chip memory usage. In addition, we also integrate NetFC into Sonata~\cite{gupta2018sonata} to detect and defense SlowLoris attack in real time.
\end{itemize}

The remainder of this paper is organized as follows. Section~\ref{sec:background} describes the background and the motivation. We present the design of our NetFC and its fundamental theory in Section~\ref{sec:design}. Section~\ref{sec:imple} shows the implementation details and our optimization mechanism. We evaluate our NetFC in Section~\ref{sec:evaluation} and present an use case that are equipped with our NetFC in Section~\ref{sec:usecase}. We finally discuss the related work in Section~\ref{sec:related} and conclude our work in Section~\ref{sec:conclusion}.

%% file: motivation/motivation.tex
\section{Background and Motivation}\label{sec:background}
In this section, we first briefly describe in-network computation, and then introduce details of floating-point standards. At last, we give two typical examples to illustrate the limitation of prior arts and motivate our work.

\subsection{Emerging Trends of In-network computation}
In-network computation, built on top of programmable switches, exploits the computational capacity of the switches to offload part of computational tasks from the server side to the network. Barefoot Networks' Tofino switches~\cite{tofino} are one of the popular programmable switches that have been widely used in academy and industry. The chip of the Tofino switch has a flexible parser and a customized match-action forwarding engine. With the provided programming language and interfaces, network programmers are able to dynamically configure the switch to program the network. Tofino switches have two multi-stage pipelines: ingress pipelines and egress pipelines. Each pipeline stage has a fixed amount of time to process packets in memory (TCAM and SRAM). The switches also support some boolean and simple arithmetic operations (e.g. integer addition and subtraction) using a set of ALUs. That said, the switches do not support complex operations (e.g. multiplication and division) or data type (e.g. floating-point number).

In-network computation is appealing for several reasons: \romannumeral1) many packets can be consumed and processed during data transmission, which significantly reduces the overhead of the network (e.g. the network queuing latency and I/O overhead); \romannumeral2) the computational workloads that are offloaded to the network can alleviate the burden of the server CPUs. 
For example, ATP~\cite{lao2021atp}, SwitchML~\cite{sapio2021scaling} and iSwitch~\cite{li2019accelerating} target at accelerating distributed deep learning via in-network gradient aggregation; NetCache~\cite{jin2017netcache} caches data in the network; NetSHa~\cite{peng2021netsha} accelerates LSH-based distributed search. We also see some traditional network algorithms~\cite{jepsen2019fast} (e.g. string matching) and network telemetry tasks~\cite{zhao2021lightguardian,yang2018elastic} (e.g. sketch) are deployed to the network. 


\subsection{Floating-point Arithmetic}\label{subsection:motivation_float_arithmetic}
\begin{figure}[!ht]
\centering
	\begin{minipage}{0.9\linewidth}
	  \centering
	  \includegraphics[scale=0.3]{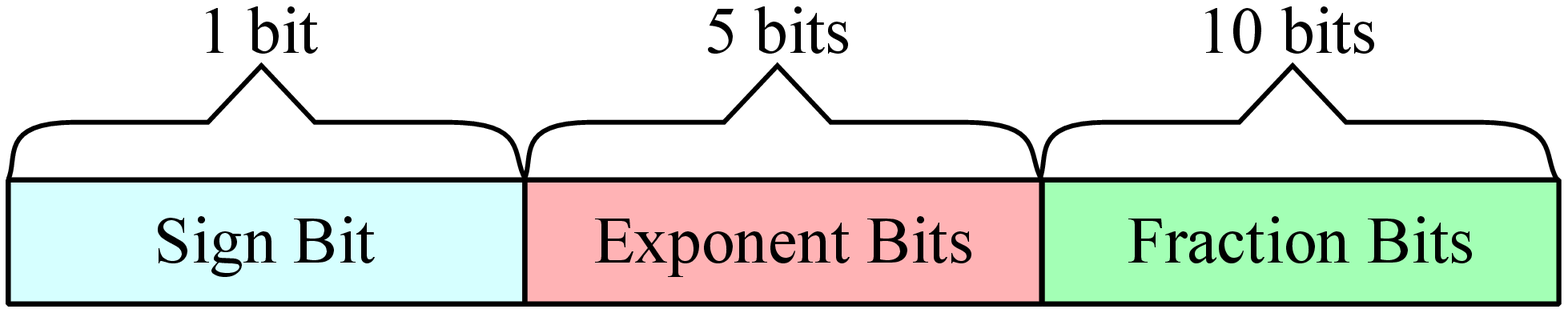}
	  \centerline{(a) 16-bit float.}
	  \hspace{2mm}
	\end{minipage}
	
	\begin{minipage}{0.9\linewidth}
	  \centering
	  \includegraphics[scale=0.3]{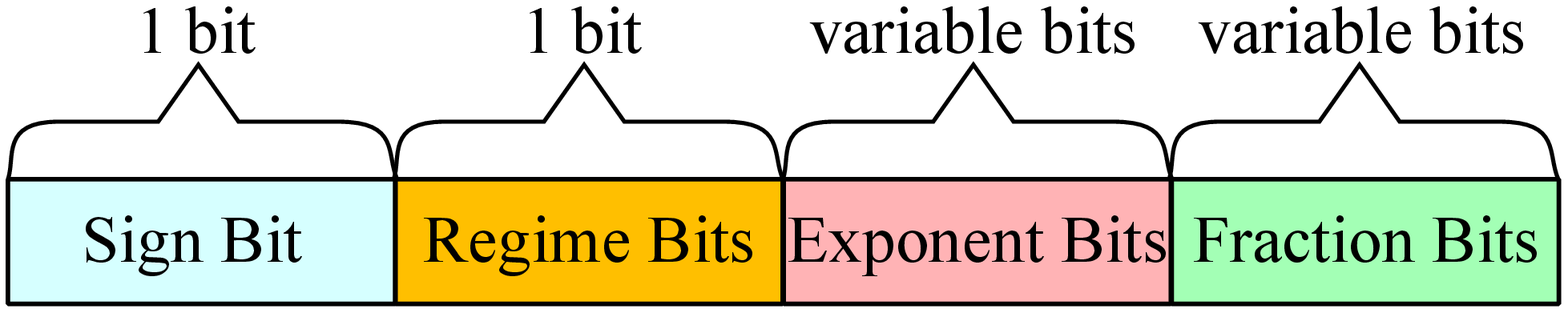}
	  \centerline{(b) 16-bit posit.}
	\end{minipage}
\caption{\small{Data format for 16-bit length floating points.}}\label{fig:data_format}
\end{figure}

In general, two popular technical standards for floating-point computation have been widely adopted: the traditional IEEE 754 standard~\cite{ieee754} and the posit standard~\cite{gustafson2017beating}. Here, we briefly introduce the two standards.

\pb{IEEE 754 floating point.} IEEE 754 float is the most widely used data type. An IEEE 754 16-bit float representation consists of three parts as shown in Figure~\ref{fig:data_format}(a): a sign bit, 5 exponent bits and 10 fraction bits. Let $e$ be the unsigned integer represented by the exponent field. If the fraction bits are {$f_1f_2...f_s$}, then $f = 1.f_1f_2...f_s$. The value $p$ of a 16-bit floating-point number that does not fall into any exception cases is given by:
\begin{equation}
p=sign*2^{e-15}* f
\end{equation}


Modern FPU (floating-point unit) implements floating-point addition, subtraction, multiplication and division as follows. 
For addition and subtraction, the FPU expresses operands with the same exponent and shift the mantissas accordingly; the shifted mantissas are then added together; for multiplication, the FPU adds the exponents of the two numbers and multiplies their mantissas; Likewise, for division, the FPU subtracts the divisor's exponent from the dividend's exponent, and divides the divisor's mantissa from the dividend's mantissa. The final outputs of the above  floating arithmetic are obtained by rounding and normalizing the results.

\pb{{Posit floating point.}} Posit is designed as a direct drop-in replacement for IEEE Standard 754 floating-point numbers (floats)~\cite{gustafson2017beating}. Figure~\ref{fig:data_format}(b) shows its data format when using 16-bit length to represent the points. Compared to float, posit data type has an additional regime field. Let $es$ be the width of regime bits ($es=1$ in Figure~\ref{fig:data_format}(b)), $k$ be the integer represented by regime field, $e$ be the unsigned integer represented by the exponent field. If the fraction bits are {$f_1f_2...f_s$}, then $ f= 1.f_1f_2...f_s$. The value of a posit number p can be represented as: 
\begin{equation}
p=sign* 2^{2^{es^{k}}}* 2^{e}* f
\end{equation}

Compared to the IEEE 754 float, posit floating point has the following advantages~\cite{gustafson2017beating,de2019posits,cococcioni2018exploiting}. 


\begin{itemize}
\item\textbf{Larger dynamic range.} Dynamic range represents a range from the minimum positive number to the maximum positive number that a number system can express. The dynamic range of 16-bit float is $6*10^{-8}$ to $7*10^4$, while the dynamic range of 16-bit posit is $4*10^{-9}$ to $3*10^8$. That said, with the same bitwidth, posit has a larger dynamic range.

\item\textbf{No representations wasted for NaN or infinity.} For the 16-bit float, when the exponent bits are 11111, it represents NaN or infinity. That said, there are 2,048 representations used to represent Nan or infinity. However, for the 16-bit posit, it represents NaN or infinity only when the representation is 0x8000.  

\item\textbf{Tapered accuracy.} Posit numbers near 1 in magnitude have more accuracy than extremely large or extremely small numbers. This phenomenon is called ``golden zone'' in~\cite{de2019posits}, in which the accuracy of posit is higher than float. For example, Posit32 has more fraction bits than Float32 numbers whose magnitude ranges from $10^{-6}$ to $10^6$. 
\end{itemize}

Due to the above characteristics, posit is considered to be very advantageous in deep learning by~\cite{langroudi2018deep,zhang2019efficient,cococcioni2018exploiting}. Indeed, the above two popular standards for floating-point calculation are widely used by different kinds of applications while our NetFC can work very well with both of them. In the following sections, we use float or floating-point (posit resp.) to refer to IEEE 754 floating point (posit floating point resp.).



\subsection{Motivating NetFC}
{
We are motivated by the fact that modern programmable switches only support simple integer arithmetic operations (addition and subtraction). That said, in-network computation tasks have to rely on other indirect `layers' to implement floating-point arithmetic operations (either addition and subtraction, or multiplication and division) if needed. These indirect `layers' may loose accuracy or increase the delay of the tasks. We take two examples here to illustrate this. 



\pb{In-network gradient aggregation.}  In-network gradient aggregation is used to accelerate distributed machine learning system~\cite{lao2021atp,sapio2021scaling}. The basic idea is to cache gradients from training workers on programmable switches, and accumulate them when some conditions are met to get aggregated gradients. In this way, the number of gradient packets sent to the parameter servers are decreased, mitigating the communication overhead. Gradients are always floating-point numbers, which require the support of floating-point arithmetic operations when performing aggregation. As programmable switches are unable to support the operations, the current solution in~\cite{lao2021atp,sapio2021scaling} is to introduce a `shim layer' on end hosts that coverts floating-point numbers to integers by multiplying a scaling factor (\emph{sf}) on workers first, and after aggregation by programmable switches (using integer arithmetic operations), the aggregated results will be forwarded to servers where they are restored back to floating-point representation by dividing \emph{sf}. The above conversion approach leads to significant accuracy loss. For example, let's consider a floating-point number $x$=1.000654 and \emph{sf}=10000. Consequently, it will convert $x$ into 10006 while the last two bits are lost.




\pb{Inband Network Telemetry.} Programmable switches are the `sweet point' to implement network telemetry as they sit in the middle of network paths. We have seen plenty of network telemetry systems built on programmable switches~\cite{gupta2018sonata, yang2018elastic,huang2018sketchlearn}. Many measurement tasks ({\eg}Slowloris attack detection) require the support of floating-point arithmetic operations (even multiplication and division). Because programmable switches cannot support these operations, these systems adopt a slow-path solution, where the intermediate results that require floating-point arithmetic operations are sent to local CPUs for processing. This solution will inevitable introduce huge delay of the tasks. Let us consider the detection of Slowloris attacks~\cite{Slowloris2009} in Sonata as a detailed example~\cite{gupta2018sonata}. The task monitors the traffic of all connections belonging to individual hosts to see whether the average traffic volume of each connection belonging to a host is less than a predefined threshold value. Apparently, this requires floating-point arithmetic operations and has to be implemented in switch local CPUs through the slow path, delaying the detection. That said, without the support of floating-point arithmetic operations in programmable switches, online detection and defense of attacks like Slowloris attacks cannot be implemented in current inband network telemetry. 

\vspace{0.5em}

\pb{Summary.} The support of floating-point arithmetic operations in programmable switches are important and essential to enable practical application of in-network computation. To the best of our knowledge, such a support is overlooked by prior arts, which motivates our work, NetFC.
}

%% file: design/design.tex
\section{Design of NetFC}\label{sec:design}
{NetFC aims at enabling sophisticated floating-point arithmetic on programmable switches nearly without accuracy loss and additional latency. In this section, we first describe the basic idea, and then discuss the challenges of NetFC, and finally detail the design.

\subsection{Design Choice}
To fix ideas, we assume that 16-bit floating-point numbers follow IEEE 754 standard, but we will relax this assumption in Section~\ref{sec:posit}. We also assume the use of Barefoot Tofino switches. Intuitively, there are two potential ways to implement floating-point arithmetic on the data plane of programmable switches.

\begin{itemize}
\item \textbf{FPU method.} A floating-point number is represented as three portions: sign, mantissa and exponent. 
For floating-point addition (subtraction reps.), it should shift the mantissa of one floating-point operand so that its exponent is identical to that of the other operand. Finally, it adds (subtracts reps.) the two operand mantissas. For floating-point multiplication (division reps.), it needs to perform multiplication (division reps.) between the two operand mantissas and addition (subtraction reps.) between the two operand exponents. 

\item \textbf{Table-lookup method.} Let's use addition to illustrate this method. For any two 16-bit floating-point operands \emph{a} and \emph{b}, we perform an addition operation between the two operands and thus get its corresponding result in advance. After this operation, we obtain a key-value pair whose key is \emph{a} and \emph{b} while \emph{z} constitutes the value. Next we traverse all possible values of \emph{a} and \emph{b} to generate multiple key-value pairs and finally constitute an addition table. Subsequently, if we calculate the sum of any other two 16-bit floating-point numbers, we only need to use the two operands as a key to look up the table while the value of the matched table entry is the result. Of course, this method can also be generalized to other arithmetic operations.
\end{itemize}

However, the question remains: \emph{can either of the above methods be deployed to programmable switches directly?} To explore this, we analyze the capacity limitations of programmable switches as follows:

\begin{enumerate}[i.]
\item \emph{Limited computation capacity.} Programmable switches (e.g. Barefoot Tofino) only supports some simple integer arithmetic. That said, floating-point numbers and arithmetic operations of multiplication and division 
have exceeded the switch capacity. 

\item \emph{Scarce on-chip memory.} Switch on-chip memory size is very small (e.g. tens of megabyte in Barefoot Tofino) so that it is impossible to provide huge memory for floating-point arithmetic. Note that, a portion of memory has to be reserved for forwarding rule storage and lookup, further aggravating this problem.

\item \emph{Limited pipeline stages.} The switch data plane often consists of a pipeline of stages, each of which is a packet processing unit equipped with some computing and storage resources. However, the number of stages is small (e.g. 32 stages in Barefoot Tofino at most), and any two dependent packet processing operations cannot be assigned to the same stage. 

\end{enumerate}

Now, let us return back to the FPU and Table-lookup methods. FPU requires on-the-fly variable shifting operations and needs to perform multiplication/division arithmetic. And thus, it is not possible to implement FPU in programmable switches. Thus we turn to the Table-lookup method.

The table-lookup method fits the programmable switches, which abstract the packet processing on data plane as tables that consist of match-action tuples. However, it does not work directly either due to a large amount of memory consumption: a 16-bit floating-point addition arithmetic would consume about 8 GB ($2^{16} \times 2^{16} \times 2B$) memory. 
Thus, the implementation finally nails down to how to reduce the memory consumption.

}

\subsection{Basic idea: Divide and Conquer}
{

\begin{figure}[!ht]
   \centering
   \includegraphics[scale=0.42]{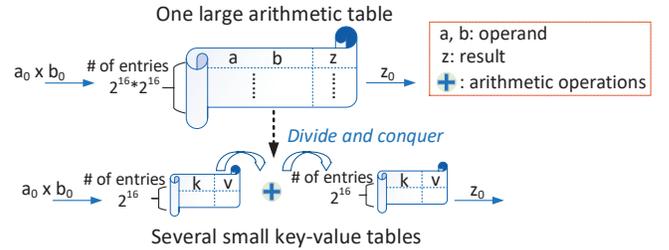}
   \caption{Example of NetFC.}\label{fig:basic}
\end{figure}

The original table-lookup method traverses all possible operands and their combinations, which finally constitutes a very large table. For example, considering two 16-bit operands, they will generate $2^{16}\times 2^{16}$ (a.k.a Cartesian product) table entries. Our NetFC adopts a divide-and-conquer approach to address this issue. Specifically, it utilizes logarithm projection and transformation to convert the original large table into several much small tables together with some simple integer arithmetic operations. We will provide its details in the subsequent sections. It is noteworthy that diverse types of floating-point arithmetics would generate different numbers of small tables. For example, as shown in Figure~\ref{fig:basic}, it uses two small tables to replace the original large arithmetic table while the total table entries are reduced from $2^{16}\times 2^{16}$ to $2^{16}$+$2^{16}$. Consequently, NetFC achieves floating-point arithmetic operations via looking up these small tables in sequence and performing some integer arithmetic operations. 

Readers might wonder looking up multiple tables and performing some arithmetic operation would degrade the performance of programmable switches. But in fact, packet-processing pipelines have an all-or-noting characteristic~\cite{gao2020switch}: programs can run at the line rate of the switch pipeline as long as they can run. Next we respectively introduce the details of NetFC for different arithmetic types.
}

{
\subsubsection{Addition and Subtraction}~\label{subsubsection:addition_design}
We assume that two floating-point numbers, $x$ and $y$, are positive; we will relax this assumption later. Note that we mainly discuss addition operation as subtraction also can viewed as one type of addition operations. Let's $i$ ($j$ resp.) denote $\lfloor \log_2(x)\rfloor$ ($\lfloor \log_2(y)\rfloor$ resp.). We note that the round-down operations would degrade the computing accuracy, and thus propose a novel approach to make up for the accuracy loss (see Section~\ref{sec:opt}). Logically, $x$ adds $y$ can be obtained as follow.

\begin{equation}
\begin{aligned}
x+y &= 2^{log_2(x+y)}\\
    &= 2^{log_2(x)+log_2(1+y/x)}\\
    &= 2^{log_2(x)+log_2(1+2^{log_2(y)-log_2(x)})}\\
    &= 2^{i+log_2(1+2^{j-i})}
\end{aligned}
\end{equation}

To achieve the above addition, we need to set up three tables (see Figure~\ref{fig:approximate}). 
The first table, \emph{logTable}, is used to record the logarithm values of all possible keys. With this basis, it is straightforward to get the value of $i$ and $j$ via looking up \emph{logTable}. The second table, \emph{miTable}, is used to figure out the value $\sigma(\theta) = \log_2(1+2^{\theta})$ for a given $\theta$; we use $j-i$ to look up \emph{miTable}, and then use the result to add $i$. Thus we can obtain the value of $i+\log_2(1+2^{j-i})$. The last table, \emph{expTable}, is to compute (find out) the exponential value for a given key. With this table, we can find out the value of $2^{i+log_2(1+2^{j-i})}$ (a.k.a $x+y$).

\begin{figure}[!ht]
   \centering
   \includegraphics[scale=0.9]{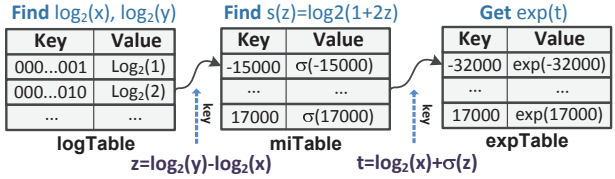}
   \caption{Floating-point arithmetic on switches.}\label{fig:approximate}
\end{figure}

Next we consider a more general condition that $x\not=0$ and $y\not=0$. Thus $x+y$ equals to $\mid x+y\mid$ or $-$$\mid x+y\mid$. Likewise, we use $i$ ($j$ resp.) to denote $\lfloor \log_2(\mid x\mid)\rfloor$ ($\lfloor \log_2(\mid y\mid)\rfloor$ resp.). $x+y$ can be obtained as follow. 

\begin{equation}
\begin{aligned}
x+y &= \pm2^{log_2(\mid x+y\mid)}\\
    &= \pm2^{log_2(\mid x\mid)+log_2(\mid x+y\mid/\mid x\mid)}\\
\end{aligned}
\end{equation}

$\mid$x+y$\mid$ belongs to one of the four situations: $\mid$x$\mid$+$\mid$y$\mid$, or $\mid$x$\mid$-$\mid$y$\mid$, or $\mid$y$\mid$-$\mid$x$\mid$, or -$\mid$x$\mid$-$\mid$y$\mid$. Finally, we can get Eq.~\ref{form:base}. 

\begin{equation}
\begin{aligned}
x+y= \pm2^{i+log(\pm1\pm2^{j-i})}\label{form:base}
\end{aligned}
\end{equation}

Indeed, due to $\pm$, Eq.~\ref{form:base} has eight possible situations, which are decided by the following three conditions: 1) x$>$0; 2) y$>$0; 3) $\mid$x$\mid$$>$$\mid$y$\mid$. The detail is shown in Table~\ref{tab:rpc}. 

\begin{table}[ht]
\centering
\caption{\small{Decision table.}}
\begin{tabular}{|c|c|c|c|}
\hline
x$>$0 & y$>$0 & $\mid$x$\mid$$>$$\mid$y$\mid$ & formula\\
\hline
T & T & T & $2^{i+log(1+2^{j-i})}$ \\
\hline
T & T & F & $2^{i+log(1+2^{j-i})}$ \\
\hline
T & F & T & $2^{i+log(1-2^{j-i})}$ \\
\hline
T & F & F & $-2^{i+log(-1+2^{j-i})}$ \\
\hline
F & T & T & $-2^{i+log(1-2^{j-i})}$ \\
\hline
F & T & F & $2^{i+log(-1+2^{j-i})}$ \\
\hline
F & F & T & $-2^{i+log(1+2^{j-i})}$ \\
\hline
F & F & F & $-2^{i+log(1+2^{j-i})}$ \\
\hline
\end{tabular}
\label{tab:rpc}
\end{table}

Similarly, we still need \emph{logTable}, \emph{miTable} and \emph{expTable}. Three variants of \emph{miTable} (i.e. $\sigma(\theta) = \log_2(1+2^{\theta})$, $\sigma(\theta) = \log_2(1-2^{\theta})$ and $\sigma(\theta) = \log_2(-1+2^{\theta})$) will be generated. In summary, NetFC maintains one \emph{logTable} table, three \emph{miTable} tables and one \emph{expTable} table to achieve floating-point addition operation.

\pb{Corner cases.} There are some corner cases, however. For example, if $x$ ($y$ resp.) equals to 0, NetFC will return $y$ ($x$ resp.) directly. In addition, in the case of $j-i>15$ ($j-i<-15$ resp.),  $x$ ($y$ resp.) is 15 orders larger than $y$ ($x$ resp.) so that the sum of $x$ and $y$ approximately equals to $x$ ($y$ resp.). 

\begin{algorithm}
    \caption{In-network floating-point addition.} 
    \begin{algorithmic}[1]
        \REQUIRE p, an input data packet. 
        \STATE parser floating-point operands $x$, $y$ from p.
        \IF {$x$ (or $y$) $\equiv$ 0}
            \STATE return $y$ (or $x$)
        \ENDIF
        \STATE get $i={\lfloor log_2(\mid x\mid)\rfloor}$, $j={\lfloor log_2(\mid y\mid)\rfloor}$ by \emph{logTable}.
        \STATE $n = j-i$.
        \IF {$n>$15}
            \STATE{\qquad}return $y$
        \ELSIF{$n<$-15}{}
            \STATE{\qquad}return $x$
        \ELSE
        \STATE {\qquad}select \emph{miTable} based on table~\ref{tab:rpc}.
        \STATE {\qquad}get $m=\lfloor log(\pm1\pm2^{n})\rfloor$ by looking up \emph{miTable}.
        \STATE {\qquad}$k=i+m$.
        \STATE {\qquad}get $\mid x+y\mid=2^k$ by looking up \emph{expTable}.
        \STATE {\qquad}set sign bit according to table~\ref{tab:rpc}.
        \ENDIF
    \end{algorithmic}\label{algo:add}
\end{algorithm}

Algorithm~\ref{algo:add} summarizes how NetFC performs floating-point addition/subtraction operations on programmable switches. First, it parses an incoming packet to obtain two operands $x$ and $y$ (line 1), checks their values (line 2 to 4) and projects them into logarithm space via looking up \emph{logTable} (line 5). After processing corer cases, it decides which \emph{miTable} to use based on Table~\ref{tab:rpc}. Finally, it further looks up the selected \emph{miTable} and \emph{expTable} to calculate the result (line 13 to line 16). Figure~\ref{fig:architecture_of_NetFC_addition} shows the implementation on data plane of programmable switches for addition/subtraction operations. 

\begin{figure}[!ht]
   \centering
   \includegraphics[scale=0.8]{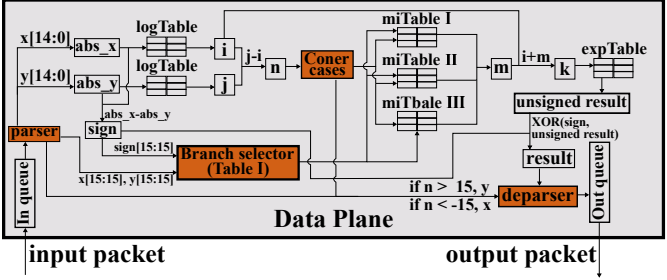}
   \caption{Implementation of floating-point addition in NetFC.}\label{fig:architecture_of_NetFC_addition}
\end{figure}
}

{\subsubsection{Multiplication and Division}
Consider two non-zero floating-point numbers, $x$ and $y$. We still use $i$ and $j$ to denote $\lfloor log_2(\mid x\mid) \rfloor$ and $\lfloor log_2(\mid y\mid) \rfloor$ respectively. Note that $x$=-$2^{\log_2(\mid x\mid)}$=-$2^{i}$ ($x$=$2^{i}$ resp.) when $x<0$ ($x>0$ resp.). Likewise, $y$=-$2^{j}$ ($y$=$2^{j}$ resp.) when $y<0$ ($y>0$ resp.). Thus the multiplication of $x$ and $y$ can be transformed to $x*y$=$\pm2^{i+j}$ while their division is $x/y$=$\pm2^{i-j}$. 

\begin{figure}[!ht]
   \centering
   \includegraphics[scale=0.54]{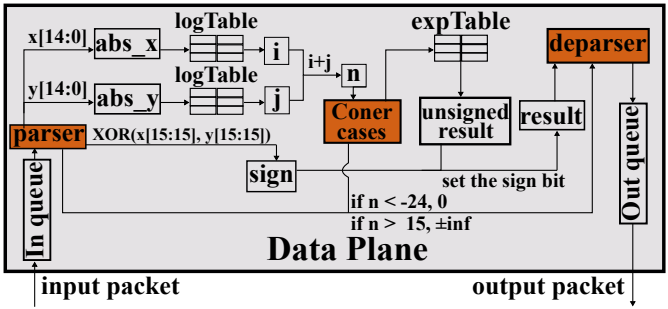}
   \caption{Implementation of floating-point multiplication in NetFC.
}\label{fig:architecture_of_NetFC_multiplication}
\end{figure}

In summary NetFC only needs two types of tables, \emph{logTable} and \emph{expTable}, to achieve floating-point multiplication and division operations. 

\pb{Corner cases.} Similarly, some corner cases should be dealt with separately. One case is that one or both operands equal to 0 so that NetFC returns 0 or NaN\footnote{NaN (Not a number) is a representation of exceptions in IEEE 754.} directly. For another case, the result exceeds the representation range of IEEE 754 floating point and NetFC would return INFINITY\footnote{All exponent bits are assigned to 1, and all fraction bits are assigned to 0.} or 0 directly.

\begin{algorithm}
    \caption{In-network floating-point multiplication.}
    \begin{algorithmic}[1]
        \REQUIRE p, an input data packet. 
        \STATE parser floating-point number $x$, $y$ from p.
        \IF {$x\equiv$0 or $y\equiv$0} {
            \STATE{\qquad}return 0.
        }
        \ENDIF
        \STATE get $i=\lfloor log(\mid x \mid)\rfloor$, $j=\lfloor log(\mid y \mid)\rfloor$ by \emph{logTable}.
        \STATE sign($x$*$y$)=XOR(sign($x$), sign($y$)).
        \STATE $n=i+j$.
        \IF {$n>$15}
            \STATE{\qquad}return INFINITY
        \ELSIF{$n<$-24}
            \STATE{\qquad}return 0.
        \ELSE
        \STATE {\qquad}get $\mid x*y \mid=2^{n}$ by looking up \emph{expTable}. 
        \STATE {\qquad}set the sign bit 
        \ENDIF
    \end{algorithmic}
    \label{algo:multiplication2}
\end{algorithm}

Algorithm~\ref{algo:multiplication2} shows how NetFC performs the floating-point multiplication. NetFC first parses an input packet to obtain two operands $x$ and $y$ and decides whether operands equal to 0 or not (line 2 to 4). Then it looks up \emph{logTable} to find out the value of $i$ and $j$, whose sum is used to detect the corner cases (line 5 to 11). After processing the corner cases, NetFC further uses the sum ($i$+$j$) to look up \emph{expTable} table and decides the sign of the result (line 13 to 14). The processing logic of floating-point division is similar to Algorithm~\ref{algo:multiplication2}. Their major differences lie in the corner cases and line 7 in Algorithm~\ref{algo:multiplication2} (\ie $n=i-j$). Figure~\ref{fig:architecture_of_NetFC_multiplication} shows the implementation of multiplication and division for floating-point numbers on programmable switches.

\subsection{Overhead analysis}
NetFC requires several tables on the switch data plane to implement floating-point arithmetic operations. This does consume the on-chip memory of programmable switches. We next discuss its overhead.

NetFC generates 5 tables at most: two \emph{logTable}\footnote{In practice, each operand needs to look up an exclusive \emph{logTable}.} (2*$2^{15}$ 16-bit entries in total, $\approx$128KB), three \emph{miTable} ($2^{15}$ 16-bit entries, $\approx$192KB) and one \emph{expTable} ($2^{16}$ 16-bit entries, $\approx$128KB). As a result, NetFC consumes about 448KB on-chip memory in total, which is reasonable considering that our lowe-end Barefoot Tofino switches are equipped with 20MB memory. Nevertheless, we propose an optimization approach to further reduce the memory usage (see Section~\ref{sec:imple}). As to pipeline usage, our experiments show that NetFC only consumes 5 pipeline stages so that it easily runs on the data plane of switches.

\subsection{Working with Posit}\label{sec:posit}
Till now we build NetFC for IEEE 754 standard, next we discuss the support for posit. Overall, there are two major differences between IEEE 754 float and posit floating point when implemented in NetFC as follows: 1) To guarantee both the input and output data are posit, the keys of \emph{logTable} and the values of \emph{expTable} need to be converted to posit floating points. 2) Since posit has larger dynamic range than that of IEEE 754 float (discussed in section~\ref{subsection:motivation_float_arithmetic}), NetFC has to consider different corner cases. For example, for float addition, the corner case is ${\mid}j-i{\mid}>15$ (discussed in section~\ref{subsubsection:addition_design}). But for posit addition, the corner case becomes ${\mid}j-i{\mid}>64$. Overall, the implementation of posit is not very different from float in principle. We believe that our NetFC is universal and it can be  generalized to other data types. 


}

%% file: implementation/implementation.tex
{
\section{Implementation and Optimization}\label{sec:imple}
In this section, we first present implementation details on programmable switches, and then introduce a few optimizations to improve the computational accuracy and reduce the overhead.

\subsection{Implementation}
We implement our NetFC on a Barefoot Tofino switch (3.2Tb/s) using $P4_{16}$ language. The switch has some restrictions on the resource usage. A particular restriction is that it cannot support complicated programming logics due to the limited pipeline stages. However, NetFC requires some \emph{if-else} conditions to decide \emph{miTable}. To address this issue, NetFC uses a separated table with pre-issued entries that covers all cases shown in Table~\ref{tab:rpc} to choose which \emph{miTable} it should use. This eliminates multi-layer nested \emph{if-else} conditions and reduces the usage of pipeline stages.

\subsection{Optimization: scaling factor}\label{sec:opt}
As mentioned before, NetFC uses $\lfloor log_2(x) \rfloor$ to approximate $log_2(x)$, which inevitably incurs accuracy loss since the decimal fraction of $log_2(x)$ has been ignored. To cope with this problem, NetFC utilizes a scaling factor, $k$, to multiply $log_2(x)$ for amplifying its decimal fraction and avoiding being ignored. NetFC also divides this scaling factor in subsequent steps for guaranteeing the correctness of floating-point arithmetic operation. 

To illustrate this, let us consider two operands, $x$ and $y$. We first get $i =\lfloor log(x)*k \rfloor$ and $j =\lfloor log(y)*k \rfloor$ by looking up \emph{logTable} respectively, and then use $\theta=i-j$ to look up \emph{miTable} for obtaining 	$\gamma=\lfloor log(\pm1\pm2^{\frac{\theta}{k}})*k \rfloor$. Finally, it looks up \emph{expTable} for the result $\lfloor \pm2^{\frac{i+\gamma}{k}} \rfloor$. It is clear that one scaling down operation (diving $k$) follows every scaling up operation (multiplying $k$). A larger $k$ brings higher accuracy, but also consumes more table entries (i.e. memory). Thus this is a tradeoff between accuracy and memory, which will be evaluated in Section~\ref{sec:evaluation}. 

\subsection{Optimization: Prefix-Based Lossless Compression}\label{sec:opt2}
We next discuss the optimization for NetFC memory overhead. Specifically, for a table in NetFC, it is possible that many continuous entries have the same value, consequently their corresponding keys can be merged. Thus we propose a prefix-based compression mechanism.

\begin{figure}[!ht]
   \centering
   \includegraphics[scale=0.30]{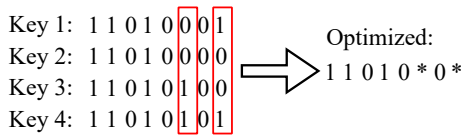}
   \caption{Example of the prefix-based compression mechanism.}~\label{fig:mask_based_compression_example}
\end{figure}

Figure~\ref{fig:mask_based_compression_example} shows an example that NetFC compresses four keys (i.e. table entry) into one key via a wildcard representation, which can be loaded to the switch TCAM memory. It is noteworthy that such compression method is lossless. Our experimental results show that it can save about 25\% memory consumption.

}

%% file: evaluation/evaluation.tex
{
\section{Evaluation}\label{sec:evaluation}
We raise questions about the computational accuracy and the overhead of NetFC:
\begin{itemize}
  \item Q1: In terms of floating-point addition/subtraction, how does NetFC perform comparing to the state-of-the-art approach (i.e. float-to-integer~\cite{lao2021atp}) (Section~\ref{sec:add_sub})?
  \item Q2: How does NetFC perform for floating-point multiplication/division (Section~\ref{sec:mul_div})?
  \item Q3: How does the scaling factor affect NetFC ( Section~\ref{sec:factor})?
  \item Q4: Does NetFC work well for the posit standard ( Section~\ref{sec:posit2})?
\end{itemize}

\subsection{Methodology}
\pb{Experimental setup.}\ We evaluate NetFC on a testbed with two commodity servers, each of which is equipped with 32 cores of Intel(R) Xeon(R) E5-2682 CPU @ 2.5GHz, 256GB RAM with Ubuntu 16.04 and Linux kernel 4.15.0-132. All servers are directly connected to a Barefoot Tofino switch (3.2 Tbps). We run NetFC on the programmable switch, and deploy a ``sender'' on one server and a ``receiver'' on the other server. The sender reads datasets and constructs floating-point operands and operators, which constitute NetFC packets to be forwarded to the switch. And the switch identifies NetFC packets and performs floating-point arithmetic operations while the receiver receives, parses and checks the results from the switch.

\pb{Baseline.}\ To the best of our knowledge, the state-of-the-art approach that perform floating-point arithmetic on the switch data plane is float-to-integer method, which has been widely used in accelerating distributed machine learning system~\cite{lao2021atp,sapio2021scaling}. However, this method only supports floating-point addition and subtraction. Thus we use it as the baseline when evaluating NetFC on addition and subtraction. But for floating-point multiplication and division, we compare NetFC with the actual results (e.g. calculated by CPUs). 

\pb{Benchmarks.}\ In general, we use two random (synthetic) datasets and one real dataset to evaluate our NetFC. One random dataset, denoted by Dataset \uppercase\expandafter{\romannumeral1}, consists of ten thousand randomly generated 16-bit floating-point numbers; the other random dataset, denoted by Dataset \uppercase\expandafter{\romannumeral2} contains ten thousand randomly generated 16-bit floating-point decimals. The real dataset, denoted by Dataset \uppercase\expandafter{\romannumeral3}, makes up of gradient updates (fifty thousand records) from a real distributed deep learning model training~\cite{jiang2019xdl}.

\pb{Metrics.}\ Let $\otimes$ denote +, -, $\times$ and $\div$. Overall, We utilize the following formula to quantize the $accuracy$:
\begin{equation}
\begin{aligned}
expect\_result &= x\otimes y \\
result &= \text{NetFC}(x\otimes y)  \\
accuracy &= e^{-\frac{{\mid}expect\_result-result{\mid}}{{\mid}expect\_result{\mid}}}     \\
\end{aligned}\label{equ:metrics}
\end{equation}
\noindent where $expect\_result$ is the exact results that are calculated by the receiver CPUs, and $accuracy$ represents the proportion of error to $expect\_result$. Thus $accuracy$ lies in between 0 and 1, a higher $accuracy$ indicates that $result$ is more close to $expect\_result$. To demonstrate this point, we assume that $accuracy$ is close enough to 1. Consequently, we can see that $\mid expect\_result-result{\mid}\approx$ $(1-accuracy)$*${\mid}expect\_result{\mid}$. This conclusion is easily to be proved via Taylor series~\cite{Taylorseries}. Our experiments show that this approximation holds as long as $accuracy$ is larger than $0.95$. 

Furthermore, we also use MSE (Mean Square Error) to measure the deviation of $expect\_result$ and $result$: 
\begin{equation}
\begin{aligned}
MSE = \frac{1}{n}\sum_{i=1}^n (expect\_result_i-result_i)^2\\
\end{aligned}
\end{equation}

\noindent where $n$ represents the total number of pairs in the dataset. MSE measures the average squared difference between $expect\_result$ and $result$, and smaller MSE values means better accuracy.

\begin{table*}[htbp]
\caption{Accuracy Table}
\begin{center}
\begin{tabular}{|c|c|p{20mm}<{\centering}|p{25mm}<{\centering}|p{15mm}<{\centering}|p{15mm}<{\centering}|p{15mm}<{\centering}|p{15mm}<{\centering}|p{15mm}<{\centering}|}
\hline
&\textbf{dataset}&\multicolumn{4}{|c|}{\textbf{IEEE 754 floating point}}
&\multicolumn{3}{|c|}{\textbf{posit floating point}}\\
\cline{3-9} 
&\textbf{name}&+,- (NetFC) &+,- (Float-to-integer)&${\times}$&${\div}$&+,-&${\times}$&${\div}$ \\
\hline
average &\uppercase\expandafter{\romannumeral1}  &99.94 &48.19 &99.96 &99.96 &99.86 &99.91 &99.91 \\
\cline{2-9} 
accuracy &\uppercase\expandafter{\romannumeral2} &99.95 &84.83 &99.94 &99.98 &99.86 &99.87 &99.95 \\
\cline{2-9} 
(\%)&\uppercase\expandafter{\romannumeral3}      &99.96 &95.28 &99.97 &99.98 & -    & -    & -\\
\hline
median &\uppercase\expandafter{\romannumeral1}   &99.94 &36.79 &100   &100   &99.91 &99.94 &99.94 \\
\cline{2-9} 
accuracy &\uppercase\expandafter{\romannumeral2} &100   &99.93 &100   &100   &99.94 &99.87 &99.96\\
\cline{2-9} 
(\%)&\uppercase\expandafter{\romannumeral3}      &100   &99.24 &100   &100   &  -   & -    & -\\
\hline
minimum &\uppercase\expandafter{\romannumeral1}  &77.88 &13.72 &60.65 &95.56 &36.79 &93.11 &89.48\\
\cline{2-9} 
accuracy &\uppercase\expandafter{\romannumeral2} &68.55 &4.54  &71.65 &99.85 &36.79 &93.11 &98.20\\
\cline{2-9} 
(\%)&\uppercase\expandafter{\romannumeral3}      &76.74 &0.01  &36.79 &99.88 &  -   & -    & -\\
\hline
\end{tabular}
\end{center}
\label{tab:accuracy_table}
\end{table*}
}

\subsection{Addition/Subtraction Performance}\label{sec:add_sub}

\begin{figure}[!ht]
\centering
\begin{minipage}{\linewidth}
  \centering
  \includegraphics[scale=1.35]{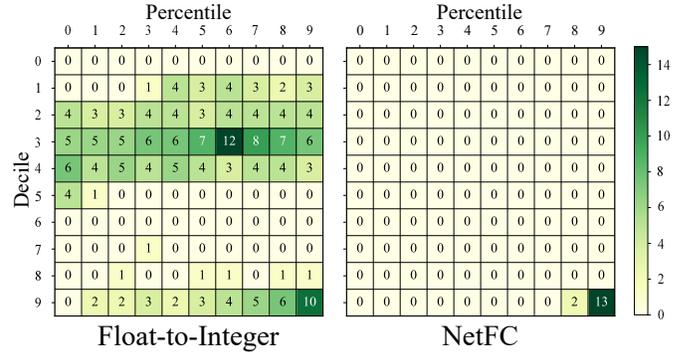}
  \centerline{(a) Dataset \uppercase\expandafter{\romannumeral1} accuracy heatmap.}
  \hspace{2mm}
\end{minipage}

\begin{minipage}{\linewidth}
  \centering
  \includegraphics[scale=1.35]{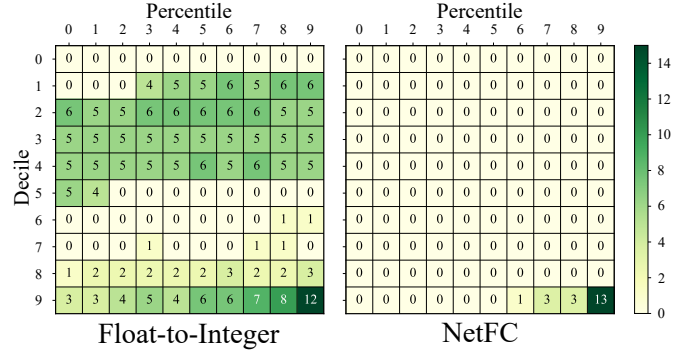}
  \centerline{(b) Dataset \uppercase\expandafter{\romannumeral2} accuracy heatmap.}
  \hspace{2mm}
\end{minipage}

\begin{minipage}{\linewidth}
  \centering
  \includegraphics[scale=1.35]{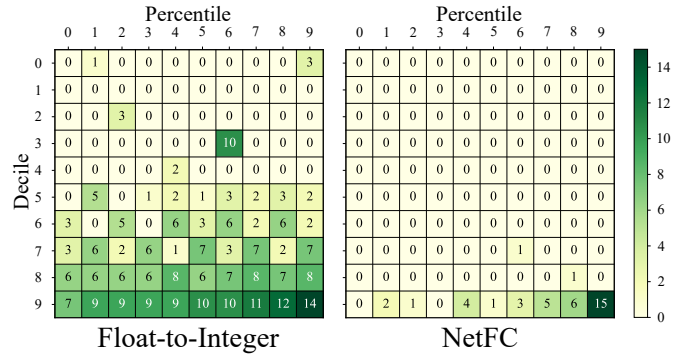}
  \centerline{(c) Dataset \uppercase\expandafter{\romannumeral3} accuracy heatmap.}
\end{minipage}
\caption{\small{Floating-point addition/subtraction heatmap. (Note that we adopt logarithm for the values in grids.) }}\label{fig:float_addtion_heatmap}
\end{figure}

For each of the three datasets, we plotted a heatmap to compare the accuracy of NetFC and Float-to-integer method. The horizontal axis represents the percentile of accuracy, and the vertical axis represents the decile of accuracy. When the value of vertical axis is $i$ and the value of horizontal axis is $j$, the value in the corresponding grid means the logarithm of the number of data with accuracy between $0.1*i+0.01*j$ and $0.1*i+0.01*(j+1)$. For example, let's consider the case that the value of axis (6,3) is 12 (see figure \ref{fig:float_addtion_heatmap}(a) left). It means that the number of data with accuracy between 0.36 and 0.37 is about 4,096 ($2^{12}$). 

As shown in Figure \ref{fig:float_addtion_heatmap}(a), the accuracy of Float-to-integer method mainly distributed near (6,3), while accuracy of NetFC concentrates around (9,9). Overflow is the main reason for the poor accuracy of Float-to-integer method. Overflow is a phenomenon that an arithmetic operation attempts to create a numeric value that is outside the range that can be represented with a given number of bits. For example, considering two 16-bits floating-point numbers $x=4.765625$ and $y = -0.005203$, if we use a scaling factor of 10,000, Float-to-integer method converts $x$ to 47656 and y to $-52$ firstly. Unfortunately, 47656 is out of the range of  16-bits integers, it will be mistaken for -17880 by the switch. This example reveals drawback (significant errors) of the Float-to-integer method: overflow occurs when the factor is large, while loss of decimal parts occurs when the factor is small. Due to this property, Float-to-integer method performs well only on decimal datasets. 

Figure \ref{fig:float_addtion_heatmap}(b) shows the performance of the two methods on the dataset \uppercase\expandafter{\romannumeral2}. As shown in figure \ref{fig:float_addtion_heatmap}(b) left, the accuracy of Float-to-integer method were concentrated in the range of 0.97 to 1.0. The accuracy is much better than the result in figure \ref{fig:float_addtion_heatmap}(a), but still not as good as NetFC (see figure \ref{fig:float_addtion_heatmap}(b) right). For NetFC, we found that only 20 out of 10,000 data have an accuracy of less than 99\%. 

In Figure \ref{fig:float_addtion_heatmap}(c), we compare the performance of the two methods on the dataset sampled from a distributed machine learning system. Again, NetFC achieves better accuracy. It is noteworthy that this dataset takes values in the range of -0.01 to 0.01, which is detrimental to our approach because the non-decimal part of the implementation is completely wasted. Thus we can remove those table entries whose value is larger than 1 since it is impossible that they would be matched. With this basis, NetFC can save more on-chip memory for further improving computational accuracy via increasing scaling factor.  

These results demonstrate the advantage of NetFC over Float-to-integer method in accuracy. Table~\ref{tab:mse} compares the two methods in terms of MSE. NetFC performs well on all three datasets;  Float-to-integer method is acceptable on dataset \uppercase\expandafter{\romannumeral3}, but performs poorly on dataset \uppercase\expandafter{\romannumeral1} and \uppercase\expandafter{\romannumeral2}.
\begin{table}[htbp]
\caption{MSE Table}
\begin{center}
\begin{tabular}{|p{24mm}<{\centering}|p{15mm}<{\centering}|p{13mm}<{\centering}|p{18mm}<{\centering}|}
\hline
  &\textbf{Dataset \uppercase\expandafter{\romannumeral1}}& \textbf{Dataset \uppercase\expandafter{\romannumeral2}}& \textbf{Dataset \uppercase\expandafter{\romannumeral3}}\\
\hline
+,- (NetFC)           & 91.32        & 2.60$\times 10^{-8}$ &1.95 $\times 10^{-13}$\\
\hline
+,- (Float-to-integer) & 198106067.16 & 0.11                 &4.17 $\times 10^{-11}$\\
\hline
${\times}$            & 28.13        & 1.52$\times 10^{-9}$ &2.72 $\times 10^{-16}$\\
\hline
${\div}$              & 27.29        & 0.98                 &1.12 $\times 10^{-4}$\\
\hline
\end{tabular}
\end{center}
\label{tab:mse}
\end{table}

\subsection{Multiplication/Division Performance}\label{sec:mul_div}


\begin{figure}[!ht]
   \centering
   \includegraphics[scale=1.35]{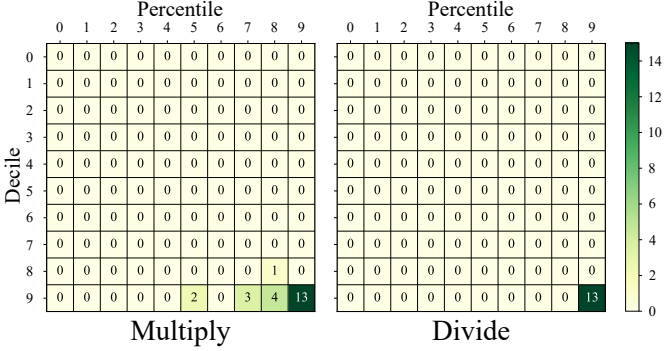}
   \caption{\small{Dataset \uppercase\expandafter{\romannumeral2} accuracy heatmap for multiplication and division.}}\label{fig:multiply_and_divide_decimal_heatmap}
\end{figure}

Figure~\ref{fig:multiply_and_divide_decimal_heatmap} shows the accuracy distribution of NetFC for multiplication and division using dataset \uppercase\expandafter{\romannumeral2}. Note that some similar results were found on the other two datasets.
We see a high accuracy no matter which datasets are used: almost all computations achieve a accuracy over 99\% for division, and only 50 (out of 10,000) computations has a accuracy of less than 99\%. Table \uppercase\expandafter{\romannumeral3} demonstrates that the deviation from $expect\_result$ of NetFC's result is very low. Table~\ref{tab:accuracy_table} shows the median accuracy of NetFC, whose worst case also can achieve up to 100\%.  


\subsection{Sensitiveness to Scaling Factor}\label{sec:factor}
\begin{figure}[!ht]

  \centering
  \includegraphics[scale=0.6]{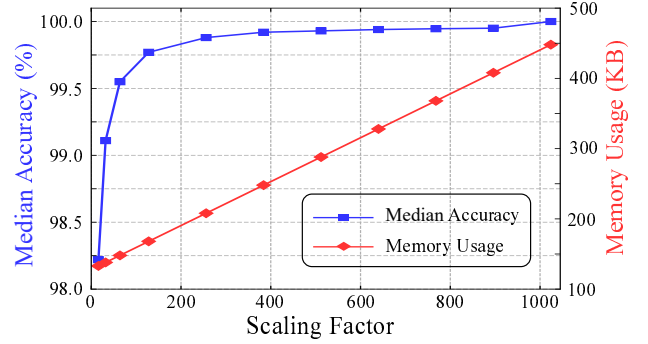}
\caption{\small{Relation between median accuracy and memory usage for floating-point addition.}}\label{fig:median_acc_and_mem_usage_addition}
\end{figure}

Recall the scaling factor is used to compensate for using round-down operation on accuracy: a larger scaling factor increases the accuracy but also the memory consumption.
Figure~\ref{fig:median_acc_and_mem_usage_addition} shows the memory usage and median accuracy when increasing scaling factor for floating-point addition. We can see that memory usage is proportional to the value of scaling factor. The accuracy increases significantly with the scaling factor when the factor below 256; beyond this point the improvement become marginal. That said, NetFC can achieve an significant accuracy improvement using limited extra cost.


\subsection{Posit Performance}\label{sec:posit2}
Table~\ref{tab:accuracy_table} shows the accuracy of NetFC for floating points of posit standard (scaling factor=512). Again, we see very high accuracy with an average accuracy above 99.86\% and a median accuracy above 99.87\%, for all four types of operations. 
These results prove NetFC's good support of floating point arithmetic operations for both IEEE 754 standard and the posit standard. 

We see a slightly lower accuracy for posit standard than IEEE 754 standard in Table~\ref{tab:accuracy_table}. The reason is that posit standard has a larger dynamic range and thus, with the same amount of memory as IEEE 754 standard, it has to use a smaller scaling factor, which is 512 (1,024 for IEEE 754 standard in our experiments). A smaller scaling factor leads to the loss in accuracy. 

%% file: usecase/usecase.tex
\section{Use Case: Online detection of Slowloris attacks}\label{sec:usecase}


\begin{figure}[htbp]
   \centering
   \includegraphics[scale=0.60]{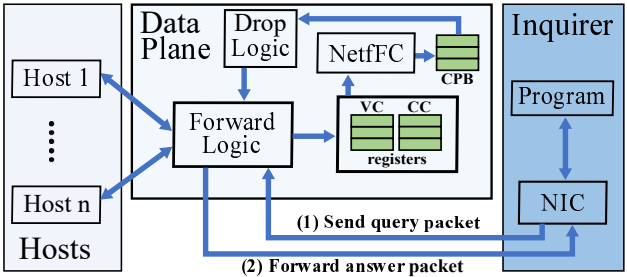}
   \caption{\small{Slowloris attack detection measurement framework with NetFC.}}~\label{fig:slowloris_with_NetFC}
\end{figure}
To verify the feasibility of NetFC in real applications, we integrated NetFC into a \emph{online} Slowloris attack detection framework, shown in Figure~\ref{fig:slowloris_with_NetFC}.

\pb{Prior Art}: Without the support of float-point multiplication and division on programmable switches, Sonata, a typical in-network telemetry system, implements Slowloris attack detection as follows: (1) Inquirer sends a query about whether there exists a slowloris attack associated with a destination IP, to data plane; (2) the system gets the number of connections and the total bytes corresponding to IP, and forward them to control plane; (3) the controller calculates the average number of connections that a byte corresponds to by dividing the number of connections with the total bytes, and then send the obtained number to the data plane, which compares the value with the predefined threshold and further answers the inquirer. This detour to local CPUs will increase the detection delay. The fundamental reason is the lack of support of division on data plane of programmable switches.

\pb{Online Slowloris attack detection with NetFC.} NetFC provides good support of floating-point division on data plane, thus enables online attack detection and defense, as shown in Figure~\ref{fig:slowloris_with_NetFC}. With NetFC, the detection no longer requires the involvement of control plane. Specially, we add a NetFC module and a CPB (connection per byte) register to the Sonata's implementation. The NetFC modules completes the division operation and writes result to CPB register.

\pb{Evaluation Setup.}\ To measure the performance of our implementation, we setup a cluster consisting five servers and an Barefoot Tofino switch. Each server has 32 CPU cores and 100G NIC, we connect them directly to switch. We use four of the five servers as host and one as inquirer. 

The experiment is divided into two phases. In the first phase, hosts communicate through switch. In this phase, when a packet arrives, our online version gets the number of corresponding connections and the total number of bytes of the destination IP, which are stored in two registers; it then computes the average number of connections each byte corresponds to through NetFC, and stores the outcome in the CPB register. 

In the second phase, the inquirer sends a query packet to the switch, the packet needs to specify the IP to be queried. The switch queries CPB register directly to get the result, and then sends the answer packet back to the inquirer. The switch can also drop the subsequent packets to this destination IP for early defence.

\pb{Results.}\ By capturing packets on the inquirer's NIC, we measured the time elapsed from the time the query packet was sent to the time we got the answer packet. Experiments show that the time to complete a query decreases from 43.156ms to 0.046ms, showing the significance that NetFC enables the removal of the involvement of switch local CPU. With such a short response time, network operators can quickly detect possible attacks
and effectively protect network from the attacks.


%% file: related/related.tex
\section{Related Work}\label{sec:related}
\noindent\pmb{FPUs.}\ FPU is a relatively conventional topic in the community. Recent researches mainly focus on how to improve the performance of floating-point calculation or increase the computational accuracy~\cite{stecklina2018lazyfp,rink2001symmetry,cretegny1998localization,oberman1996design}. However, due to a specific programming model and limited computing capacity, it is not easy to deploy FPUs into programmable switches.

\noindent\pmb{Logarithm Number System Processor.}\ Logarithm Number System (LNS) is another numeric representation. Recently, the community has proposed different approaches to optimize the LNS processors~\cite{yu199130,coleman199932,coleman2000arithmetic}. Actually, LNS mainly uses such logarithm representation to assist its microprocessors for higher computational performance. On the contrary, our NetFC borrows this representation to achieve on-the-fly floating-point operations on programmable switches. 

\noindent\pmb{In-network computation.}\ There are a lot of works focusing on in-network computation and achieving significant performance improvement. For example,~\cite{lao2021atp,sapio2021scaling} perform in-network gradient aggregation to accelerate distributed deep learning. NetCache~\cite{jin2017netcache} implements a high-performance key-value cache in programmable switches. Sonata~\cite{gupta2018sonata} utilizes programmable switches to achieve fast In-band network telemetry. However, these works do not fundamentally solve the floating-point calculation issue on the data plane of programmable switches.




%% file: conclusion/conclusion.tex
\section{Conclusion}\label{sec:conclusion}
In-network computation is an emerging trend to reduce the network overhead and accelerate data-intensive applications via offloading some computational tasks to the network (i.e. programmable switches). However, programmable switches only support simple integer arithmetic operations while other sophisticated floating-point operations have exceeded their capacity. To address this issue, the prior arts propose a float-to-integer mechanism or offload such computational tasks to the local CPUs of switches, which may either lead to accuracy loss or incur additional latency. To this end, we design and implement NetFC, a table-lookup method, to achieve on-the-fly in-network floating-point operations nearly without accuracy loss. NetFC adopts divide-and-conquer mechanism and prefix-based lossless compression method to reduce memory consumption, designs a scaling-factor approach to improve computational accuracy. 
Our experimental results show that the average accuracy of NetFC can achieve up to 99.94\% at worst with only 448KB memory consumption. Furthermore, we integrate NetFC into Sonata for detecting Slowloris attack, yielding significant decrease of detection delay. 
We believe that our NetFC can be a building block for in-network computation.

